# Popper's falsification and corroboration from the statistical perspectives

Youngjo Lee and Yudi Pawitan
Seoul National University and Karolinska Institutet
Emails: youngjo@snu.ac.kr and yudi.pawitan@ki.se

The role of probability appears unchallenged as the key measure of uncertainty, used among other things for practical induction in the empirical sciences. Yet, Popper was emphatic in his rejection of inductive probability and of the logical probability of hypotheses; furthermore, for him, the degree of corroboration cannot be a probability. Instead he proposed a deductive method of testing. In many ways this dialectic tension has many parallels in statistics, with the Bayesians on logico-inductive side vs the non-Bayesians or the frequentists on the other side. Simplistically Popper seems to be on the frequentist side, but recent synthesis on the non-Bayesian side might direct the Popperian views to a more nuanced destination. Logical probability seems perfectly suited to measure partial evidence or support, so what can we use if we are to reject it? For the past 100 years, statisticians have also developed a related concept called likelihood, which has played a central role in statistical modelling and inference. Remarkably, this Fisherian concept of uncertainty is largely unknown or at least severely under-appreciated in non-statistical literature. As a measure of corroboration, the likelihood satisfies the Popperian requirement that it is not a probability. Our aim is to introduce the likelihood and its recent extension via a discussion of two well-known logical fallacies in order to highlight that its lack of recognition may have led to unnecessary confusion in our discourse about falsification and corroboration of hypotheses. We highlight the 100 years of development of likelihood concepts. The year 2021 will mark the 100-year anniversary of the likelihood, so with this paper we wish it a long life and increased appreciation in non-statistical literature.

**Running title:** Falsification to Confirmation

# Introduction

Practical induction, including falsification and corroboration of propositions (scientific hypotheses, universal laws, social theories etc.), relies on information from limited data, hence must deal with uncertainty. Probability is the most recognized mathematical representation of uncertainty, yet Popper was emphatic in his rejection of inductive probability. Our goal is to bridge this gap. For the purpose of inductive reasoning, we can distinguish three concepts of probability. The first concept concerns the logical probability of a proposition being true as a measure of the degree of belief; this comes in two versions. The objective version (Keynes, 1921) express rational degrees of belief, but it was never adopted by mathematicians and statisticians. The subjective probability, proposed by series of papers by Ramsey in 1920s (Ramsey, 1931) and de Finetti in 1930s (de Finetti, 1974), became the dominant version, and later known as Bayesian probability. The second probability concept relates to the long-run rate of observable and repeatable events (Von Mises,1928), mathematically formalized by Kolmogorov in 1933. And the third, Popper's propensity theory of probability defines it as the property of the generating mechanism – for instance, the coin toss. Frequentist statisticians interpret probability as having both the long-run and propensity properties. In orthodox statistical inference, the probability is associated with the P-value and the confidence intervals, routine quantities produced in virtually all scientific hypothesis testing. They are Popperian in the sense of being deductive quantities, but they are often interpreted inductively as evidence from observations.

In this paper we highlight the likelihood and confidence as alternative non-probabilistic measures of uncertainty that are Popperian-compliant measures of corroboration. We shall discuss two well-known fallacies as specific illustrations. The conjunction fallacy has highlighted the limitations of probability-based reasoning as the rational basis of confirmation. The prosecutor's fallacy also arises from confusing probability and likelihood. To a large extent both fallacies vindicate Popper's objection to probability-based corroboration. We also discuss the results of a fascinating recent study that showed traces of likelihood-based reasoning in 15-month-old infants, indicating that such mode of reasoning is natural in human thinking.

In statistical terminology, anything unknown is called a parameter. It could the true speed of light; when we measure it, we only get an estimate with some margin of errors. The true lethality of covd-19 is an unknown parameter; reported values are estimates that are uncertain because of statistical errors due to bias ascertainment and sampling variability. Statistical hypotheses are statements that can be re-expressed in terms of these underlying parameters. The classical likelihood was introduced by RA Fisher in 1921 to deal with unknown fixed parameters such as the speed of light or the viral lethality. There have been many attempts to extend the concept to accommodate random unknowns, but all failed (Lee et al., 2017). In 1996, Lee and Nelder introduced extended likelihood approach for inferences of random unknowns. The confidence interval, perhaps the most important statistical quantity in scientific reports, has an unknown truth status. Whether or not a particular confidence interval covers the true parameter value is a random unknown. Theoretically it is a binary random variable, whose probability is represented by the confidence level. Confidence and likelihood have been fundamental statistical concepts with distinct technical interpretation and usage. We have shown recently (Pawitan and Lee 2020) that confidence is in fact an extended likelihood, thus giving a much closer correspondence between the two concepts.

To illustrate, let G be a general proposition, such as "All ravens are black", and E be a particular proposition or an observation (evidence) such as "The raven in front of me is black." A deductive reasoning "G implies E" can attain complete confidence, given the basic premises – such as the axioms -- are true. But, in practical induction, we want to establish G after observing E, which is logically possible but we know we may not attain complete confidence. We can use the logical probability to represent the deductive logic "G implies E" as

$$\text{Prob}(E|G) = 1 \text{ and } \text{Prob}(\text{not } E|G) = 0.$$

Probability extends classical binary logic, as the logical probability can be quantified as a number between 0 and 1 to capture a degree of belief. The extreme value 0 indicates impossibility (the proposition is false), and 1 certainty (the proposition is true). To use the likelihood for inductive inference, let $\theta$ be the probability of black ravens. Then G corresponds to $[\theta=1]$. Given the observation (evidence) E, we define the likelihood as

$$L(\theta) = P(E|\theta).$$

We immediately see how the likelihood captures the information in E. If E is a white raven, then $L(\theta=1)=0$, meaning that, on observing a non-black raven, G has zero likelihood, hence falsified. Observing E being a black raven is certain when $\theta=1$, so in this case $L(\theta=1)=1$. But it is also close to certain for $\theta$ near but less than 1, i.e. $L(\text{not } G) \sim 1$, so we cannot establish G based on observing a single black raven. But the evidence gathers strength if we go around the world observing 1000 ravens and find that they are all black. However, the likelihood function is never exactly zero for $\theta$ near but not exactly 1, indicating we are never certain.

The confidence approach is another practical implementation of Popperian views. Traditionally, given data E, construct confidence intervals with a certain probability of being true. For example, for the raven example: we want to establish G, which corresponds to $\theta=1$. If we observe 10 ravens and all are black, the 95% confidence interval is 0.69 to 1.00 (Pawitan, 2001, Chapter 5). Thus, since 1.00 is in the interval, E does not falsify G that all ravens are black; it does support G, but the support is not that strong. Even when 31% of the ravens are non-black ($\theta=0.69$), there is still a good chance to observe 10 black ravens in a sample of 10 ravens. Increasing the confidence say to 99% will only widen the interval 0.59 to 1.00. But increasing the sample size to 1000, the 95% confidence interval is now 0.996 to 1.000. So, there is a very strong support, but still not certainty, about G. Theoretically, we only get certainty with an infinite ample, but we can reach practical certainty. This simple example shows that likelihood and confidence can operationalize Popper's (1959) fallibilism for scientific development, where we can falsify, but cannot prove G; but it also clear that in practice we can build evidence for it. Traditional confidence procedure has a propensity interpretation: the 99% probability is the property of the procedure, not of the observed interval. So, it is fully Popperian. Recent developments in the confidence theory has brought the idea of confidence distribution, which is closely related to the Bayesian posterior probability. However, crucially, *confidence is still not a probability*, so it still conforms to the Popperian demand of non-probabilistic corroboration.

In summary, this simple example shows that likelihood and confidence can operationalize Popper's program for scientific development, where (i) we can falsify, but not prove G; (ii) they are not probability, so we can use them for non-probabilistic corroboration of hypotheses. Yet, there is a close logical connection between confidence and Bayesian subjective probability. In this sense, confidence bridges the gap between the inductive and Popperian non-inductive views. So we do not see a clear demarcation between these two views.

In an era of artificial intelligence, induction and learning from data becomes more crucial for drawing valid inferences. Originally, the goal of science was to prove propositions, to establish scientific theory based on observational data. However, the difficulty of such an inductive process has been recognized since the Greek and Roman periods. Hume (1748) argued that inductive reasoning cannot be justified rationally, because it presupposes that the future will resemble the past and the present. A resolution to the induction problem offered by Kant (1781) is to consider propositions as valid, absolutely a priori. Popper (1959) proposed falsification of propositions instead of proving them to be true. As Broad (1923) stated "induction is the glory of science but the scandal of philosophy". However, it could be also a scandal of science if there is no way to confirm scientific theory with the complete confidence via induction. Jeffreys (1939) provided a way of confirming scientific theories by modifying Bayesian approach but requires an infinite evidence, which is not possible in practice.

Compared with scepticism, fallibilism does not reject the existence of knowledge. However, fallibilism recognize that there are not any reliable means of justifying knowledge as true or probable. Thus, justification is the misconception that knowledge can be genuine or reliable only if it is justified by some source or criterion (Deutsch, 2011). Contemporary Bayesian confirmation theory has been developed to overcome various fallacy. The conjunction fallacy has highlighted the limitations of Bayesian probability-based reasoning as the rational basis of confirmation. We discuss the results of a fascinating recent study that showed traces of likelihood-based reasoning in 15-month-old infants, indicating that such mode of reasoning is natural in human thinking. In this article, we also show that via extended likelihood, one can obtain complete confidence (100% degree of belief) regarding a general proposition with finite evidence.

# Probability-based reasoning for inductive inference

Popper was against probability-based induction, so it is instructive to see which particular aspects were anathema to him. Induction is a form of reasoning from particular examples to the general rule, in which one infers a proposition based on evidence (data or observations). However, establishing the truth of a general proposition is problematic, because it is always possible that a conflicting observation to occur. This problem is known as the induction problem. Originally, the goal of science was to confirm general propositions (scientific theories) such as "All ravens are black", or to infer them from observational data. However, the difficulty of deriving such inductive logic has been recognized since the Greek and Roman periods.

As Cox (1946) noted, the logical probability represents the equivalence of 'G implies E' as

$$\text{Prob}(E|G)=1 \text{ (and therefore Prob(not } E|G)=0)$$

with 'not E implies not G' as

$$\text{Prob}(G|\text{not } E) = \text{Prob(not } E|G)\text{Prob}(G)/\text{Prob(not } E) = 0, \text{ and Prob(not } G|\text{not } E)=1,$$

provided that the denominator is not zero. From this, we see clearly that one observation of a non-black raven can certainly falsify the general proposition. Popper (1959) saw this falsifiability of a proposition as a useful demarcation for scientific theory. Thus, he used deductive rule for scientific theory.

Via inductive rule, we want to establish G after observing E. The Bayes rule is

$$\text{Prob}(G|E) = \text{Prob}(E|G)\text{Prob}(G)/\text{Prob}(E), \qquad (1)$$

provided $\text{Prob}(E) > 0$. Thus, if G implies E, we have

$$\text{Prob}(G|E) = \text{Prob}(G)/\text{Prob}(E) \geq \text{Prob}(G),$$

with strict inequality if $0<\text{Prob}(E)<1$, meaning that observing E increases the probability of G, unless $\text{Prob}(G)=0$ or 1. Hence, a particular observation can corroborate, but it is not clear how much the evidence contributes to an increase of logical probability of the general proposition. Kant (1781) proposed a resolution to the induction problem, which involved considering the propositions as valid, absolutely a priori, i.e. $\text{Prob}(G)=1$. He thought that Euclidean geometry is self-evidently true. In this case

$$\text{Prob}(G|E) \geq \text{Prob}(G) =1, \text{ so Prob}(G|E) =1.$$

Thus, no evidence can increase of the logical probability of G. Broad (1923) stated that "induction is the glory of science but the scandal of philosophy" by showing that

$$0 = \text{Prob}(G|E) \geq \text{Prob}(G)=0.$$

Popper (1959, Appendix vii) also explained why he believed that $\text{Prob}(G)=0$ a priori and used it as argument against probability-based induction. If we know whether G is true or false a priori, its logical probability cannot be altered by future evidence. This seems to be a key aspect of the logical probability of induction that he found unacceptable. We show below that $\text{Prob}(G)=0$ is actually not a necessary condition.

**The logical probability of Pascal**

Throughout the article we define $I(H)$ as the truth function of whether the proposition (or hypothesis) H is true or not, i.e.

$$I(H)=1 \text{ if H is true or } I(H)=0 \text{ if it is false.}$$

As the value of $I(H)$ is unknown, let us treat it as an unobservable random variable. Then, we can represent logical probability as follows:

$$\text{Prob}(E) = \text{Prob}[\ I(E)=1\ ] \text{ and}$$

$$\text{Prob}(G|E) = \text{Prob}[\ I(G)=1|I(E)=1\ ] = \text{Prob}(G)/\text{Prob}(E).$$

Thus, Prob(G|E)=Prob(G), i.e. a new evidence cannot increase the logical probability of general proposition when Prob(G) is either 0 or 1; the two extreme cases of Kant and Popper.

Pascal's wager was published posthumously in *Pensées* ("Thoughts"). Given that reason alone cannot determine whether E is true or not, i.e. $T \equiv I(E) = 1$ or 0, Pascal concluded the uncertainty associated with this question can be expressed as probability Prob(T=1), treating T as an outcome from a coin toss but without seeing it. He further insisted even if we do not know the outcome of this coin toss, we must evaluate our actions based on the expectation of the consequence caused by our action. Pascal -- one of the founders of probability theory -- then treated his logical probability as Kolmogorov's probability. Pascal's argument of probability does not involve any data, so that he might be the first who used prior probability. He did not mention how to compute his prior probability. He also did not have any reason to distinguish his logical probability from Kolmogorov's because he lived in the 17th century, so unaware of 20th century's mathematical probability. When we discuss the probabilities of Bayes (1763), Laplace (1814) and Fisher (1930), they also need not formulate the same probability as Kolmogorov's (1933) probability.

Pascal interpreted probability as logical probability of any proposition, such as Trump's impeachment. Ramsey and de Finetti interpreted it as a betting quotient and showed that, to be coherent, it should satisfy additive probability axioms. Otherwise someone can run a Dutch Book (arbitrage) on your betting. Suppose that you set your betting quotients as follows: (i) bet that the proposition E is true, risking $\alpha$ to win $1-\alpha$, and (ii) bet that E is not true, risking $\beta$ to win $1-\beta$. Then I will make *two bets* against you, on E and (not E) simultaneously, either as a player or as a bookie depending on $(\alpha + \beta)$ below. My expected winning from the two bets is

$$(1-\alpha)\ \text{Prob}(E) - \alpha\ (1-\text{Prob}(E)) + (1-\beta)\ \text{Prob}(\text{not } E) - \beta\ (1-\text{Prob}(\text{not } E))$$
$$= (1-\alpha-\beta)\ (\text{Prob}(E)+\text{Prob}(\text{not } E)) = 1-\alpha-\beta.$$

If $(\alpha + \beta) < 1$ I can always win by betting as player, and if $(\alpha + \beta) > 1$ I can always win by becoming a bookie. Thus, the betting quotients $\alpha$ and $\beta$ should satisfy probability laws such that $(\alpha + \beta) = 1$ and $\text{Prob}(E) = \alpha \geq 0$ and $\text{Prob}(\text{not } E) = \beta = 1 - \alpha \geq 0$. They call resulting logical probability subjective, because different people may have different betting quotients. According to Ramsey and de Finetti, we may know someone's personal probability via their betting behavior. Can people agree on their betting quotient? A question is whether there exists an objective logical probability. Objective Bayesians aim to have an objective logical probability by finding an ignorant prior probability, whereas we want to find it without presuming a prior.

The Bernoulli model was developed for observable binary random events such as coin tossing. In coin tossing, the true probability can be determined by long-run frequency (Von Mises, 1928), whereas in logical probability the truth or falsity of proposition can never be observable, so that the long-run frequency is not available. Furthermore, expectation is for repeatable events not for a single event such as truthfulness of certain proposition. The interpretation of logical probability can be seen as an extension of propositional logic that enables reasoning with proposition whose truth or falsity is unknown. In this article, we show how recently developed likelihood theories helps to understand logical probability. We first briefly review developments of existing Bayesian approach in the last three centuries.

An inductive logic is based on the idea that the probability represents a logical relation between the proposition and the observations. Accordingly, a theory of induction should explain how one can ascertain that certain observations establish a degree of belief (logical probability) strong enough to confirm a given proposition. The sunrise problem is a quintessential example of the induction problem, which was first introduced by Laplace (1814). However, in Laplace's solution, a zero probability was assigned to the proposition that the sun will rise forever, regardless of the number of observations made. Therefore, it has often been stated that complete confidence regarding a general proposition can never be attained via induction. We explain why such an extreme view was formed. Lee (2020) shows that through induction, one can rationally gain complete confidence in general propositions via likelihood based procedure.

**Bayesian and Fisher's logical probabilities**

Pascal would first introduce logical probability but without the data and did not show how to compute it. It was Bayes (1763) who introduced Bayesian approach to set logical prior probability and update it based on the data. However, he might not have embraced the broad application scope now known as Bayesianism, which was in fact pioneered and popularized by Laplace (1814) as an inverse probability. Bayesianism has been applied to all types of propositions in scientific and other fields (Paulos, 2011). Savage (1954) provided an axiomatic basis for the Bayesian probability as a subjective probability, whereas Jaynes (2003) provided as an objective probability. Whether it comes from objective or subjective Bayesian schools, Bayesians are common not to distinguish Kolmogorov's mathematical and logical probabilities because their axioms allow their logical probability satisfies properties of mathematical probability and require prior to form their logical probabilities. Fisher (1930) formed a logical probability without presuming a prior and axioms, so that his probability does not necessarily satisfy properties of Kolmogorov's mathematical probability even though he believed unfortunately so until his death.

**Laplace's solution to the sunrise problem**

Using the sunrise problem, Laplace (1814) demonstrated how to compute such an actual logical probability based on the data. Let θ be the long-run frequency of sunrises, i.e., the sun rises on 100×θ% of days. Under the Bernoulli model, the general proposition G that "The sun rises forever" is equivalent to the hypothesis θ = 1. The general proposition for which θ = 1 is then a Popperian scientific theory because it can be falsified if a conflicting observation, i.e., one day of no sunrise, occurs. Based on finite observations until now, could it possible allow for complete confidence on θ = 1?

To represent a description of the uncertainty about the true value of θ, prior probability should be assigned in Bayesian approach. Prior to the knowledge of any sunrise, suppose that one is completely ignorant of the value of θ. Laplace (1814) represented this prior ignorance about θ by means of a uniform prior on θ ∈ [0,1]. This uniform prior was proposed by both Bayes (1763) and Laplace (1814) as Bayes-Laplace postulate of insufficient reason. Given the value of θ and no other information relevant to the question of whether the sun will rise tomorrow, Laplace computed the probability of the particular proposition E that "The sun will rise tomorrow." Laplace, based on a young-earth creationist reading of the Bible, inferred the number of days by considering that the universe was created approximately 6000 years ago. He computed the posterior, given n = 6000×365 =102 2,190,000 days of consecutive sunrises,

Prob(E| n days of consecutive sunrises) = 0.9999995;

see Lee (2020). This probability of this particular proposition, that is, the sun rising the next day given n days of consecutive sunrises, is eventually one as the number of observations n increases. However, this aspect is not sufficient to confirm the general proposition G that the sun rises forever. Broad (1918) showed that

Prob(G| n days of consecutive sunrises) = 0 for all n.

Hence there is no justification whatsoever for attaching even a moderate probability to a general proposition if the possible instances of the rule are many times more numerous than the instances already investigated for a more thorough discussion.

Jaynes (2003) argued that a beta prior density, Beta(α,β) with α > 0 and β > 0, describes the state of knowledge that we have observed α successes and β failures prior to the experiment. The Bayes–Laplace uniform prior, Beta(1,1), means that a trustworthy manufacturer sent you a coin with information that he/she observed one head and one tail in two trials before sending the coin, i.e. Prob(G)=0. Even if you have an experiment with heads only for many trials, there is no way to attain complete confidence on heads only, unless you discard the manufacturer's information. Contemporary Jeffreys' prior (1939) and Bernardo's (1979) reference prior of objective Bayesian approach is Beta(1/2,1/2), but Lee (2020) showed that under this objective Bayesian prior P(G| n days of consecutive sunrises) = 0. Thus, the Bayes-Laplace approach, even if it is derived from objective Bayesianism, cannot overcome the degree of skepticism raised by Hume (1748) because they presume Prob(G)=0.

Let E be an event of "n consecutive days of sunrises," satisfying Prob(E|G)=1. Laplace's (1814) solution shows that Prob(G|E) = 0 for any large n, because he presumed unfortunately Prob(G)=0 a priori, which cannot be an ignorant prior. This leads to

0=Prob(G|E) ≥ Prob(G), so Prob(G)=0.

In his *Logic of Scientific Discovery,* Popper (1959) also explained why, he thinks, Prob(G)=0, so that any evidence cannot alter the logical probability of general proposition Prob(G|E)=0, which precludes probability-based induction.

**Jeffreys's resolution**

That a general proposition cannot be confirmed via scientific induction based on the Bayes–Laplace formulation turns out to be because the choice of the prior had been wrong. Jeffreys' (1939) resolution was another prior, which places a mass 1/2 on the general proposition $\theta = 1$ Prob(G)=1/2 and a uniform prior on [0,1) with 1/2 weight. Let E be an event of n days of consecutive sunrises. This leads to

$$0 < \text{Prob}(G) = 1/2 < \text{Prob}(G|\text{one day of sunrises}) = 2/3$$
$$< \text{Prob}(G|\text{ two days of consecutive sunrises}) = 3/4 \cdots$$
$$< \text{Prob}(G|E) = (n+1)/(n+2) \cdots$$

and thus, Prob(G|E) increases to one eventually as n increases (Lee, 2020). Jeffreys' resolution produces an important innovation of the Bayesian hypothesis testing (Etz and Wagenmakers, 2017). Senn (2009) considered Jeffreys' (1939) work as "a touch of genius, necessary to rescue the Laplacian formulation of induction. However, with Jeffreys's resolution, the scientific induction cannot attain complete confidence even in this era of big data, because such a process requires infinite evidence, i.e. P(G|E)=1 only when the evidence is infinite. A key of Jefferys's resolution is to presume Prob(G)=1/2 a priori. But there is another way to rescue the Laplacian formulation of induction without presumption of a prior.

**Confirmation for general proposition**

Carnap's (1950) degree of confirmation of the general proposition G by the evidence E is

$$C(G,E) = \text{Prob}(G|E) - \text{Prob}(G) \leq 1 - \text{Prob}(G) = \text{Prob}(\text{not } G).$$

Popper (1959) preferred 'corroboration' over 'confirmation', and 'testability' over 'confirmability,' since a theory is never confirmed to be true, i.e. Prob(G)=0 for him. However, the term 'confirmation' has survived in the literature, so we also use it here to mean 'corroboration.' Since under Bayes-Laplace (Kant) formulation, Prob(G|E) = Prob(G) = 0 (Prob(G|E) = Prob(G) = 1) to give

$$C(G,E) = 0.$$

In Carnap's inductive logic (1950), the degree of confirmation of every universal law is always zero. However, we see that it is because they presume Prob(G)=0. Therefore, under Prob(G)=0 a priori the universal law can never be accepted, but is not rejected until conflicting evidence appears. In Jeffreys' (1939) resolution with Prob(G) = 1/2,

$$C(G,E) = \text{Prob}(G|E) - \text{Prob}(G) = n/\{2(n+2)\} > 0,$$

and thus the evidence E confirms the general theory G positively. But how we can justify Prob(G)=1/2. However, in the confidence resolution, although the prior Prob(G) is not assumed, complete confidence (confirmation) Prob(G|E) = 1 (Prob(U=I(G)=1)=1) is achieved.

**Confidence as an alternative to Bayesian logical probability**

We see that Bayesian solution can give different conclusions, depending upon the choice of priors. Many writers have criticized the use of arbitrary priors. The question is whether we can form an objective logical probability without presupposing a prior. In 1930 Fisher showed that it is possible and called the idea "fiducial probability," which has largely been abandoned in practical statistics, but it did lead to the "confidence" concept. Confidence interval is one of the most widely used statistical inference tools in practice.

For example, in polls before election, the 95% confidence interval of the true voting rate $\theta_0$ of a certain candidate is reported. Based on poll data, suppose that the $\alpha \times 100\%$ confidence interval [L,U] is reported. Let

$$T \equiv I(L \leq \theta_0 \leq U),$$

where T= 1 if the interval covers the truth, and 0 otherwise. Thus, T (as a function of data) is a binary random variable with probability

$$\text{Prob}(T = 1) = \alpha.$$

When it *refers to the observed interval*, $\alpha$ is called the confidence, rather than probability, of the interval because it is not a proper Kolmogorov probability (Schweder and Hjort, 2016). Let E be the proposition that the specific interval [L,U] covers the true paramete $\theta_0$ and T=I(E). We shall see that the confidence is indeed a way of computing Pascal's logical probability of I(H) and an alternative to Bayesian probability without assuming a prior. The confidence statement of the interval, the true value of $\theta$ is contained in the interval, is sample dependent proposition, and we attain complete confidence when $\alpha = 1$.

Fisher's (1930) classical likelihood is for inferences about fixed unknowns. Lee and Nelder (1996) extended likelihood inferences to unobservable random variable such as T. Pawitan and Lee (2020a) showed that the confidence is the extended likelihood of Lee and Nelder (1996). For an observed confidence interval [l,u], the value T = t = I(l $\leq \theta_0 \leq$ u) is realized but still unknown single event, because the true parameter value $\theta_0$ is unknown, giving

$$L(t=1) \equiv \text{Prob}(T=1) = \alpha. \tag{2}$$

Here L(t=1) is the extended likelihood of single (realized) event; it is equal to the confidence. The confidence interval is justified theoretically in terms of its coverage probability, which is given a Popperian propensity interpretation as belonging to the procedure (Gillies, 2000). But in (2) we address the epistemic question if there is a probabilistic way to state our sense of uncertainty in an observed confidence interval. In coin tossing, we can compute the long-run frequency as a true probability. However, in the confidence concept the realized value t is unobservable, so its long run frequency is not meaningful. Instead, frequentists use coverage probability in hypothetical repetitions of constructing confidence intervals as a thought experiment. If we construct confidence interval 100 times repeatedly from the same experiments, 100 x $\alpha$ of them will cover true value of $\theta$. The coverage probability is a long-run rate of the coverage of the confidence interval in hypothetical repetitions. Thus, the confidence concept is a bridge between the Kolmogorov and logical probabilities.

The P-value has been widely used for scientific inferences. Let X be a sufficient statistics for $\theta$. Fisher (1930) derived the fiducial probability of $\theta$. Define the right-side P-value function

$$C(x, \theta) = \text{Prob}(X \geq x | \theta).$$

Given $X=x$, as a function of $\theta$, $C(x,-\infty) = 0$ and $C(x,\infty) = 1$ and $C(x, \theta)$ is a strictly increasing function of $\theta$. Thus, $C(x, \theta)$ behaves as if it is the cumulative distribution of $\theta$. This leads to the fiducial probability for $\theta$

$$c(x, \theta) = dC(x, \theta)/d\theta,$$

which is derived without presupposing a prior. Schweder and Hjort (2016) called it the confidence density. Pawitan and Lee (2020a) showed that this sample-dependent confidence is indeed an extended likelihood with updating rule

$$c[(x1,x2), \theta] \propto c(x1,\theta)L(\theta; x2),$$

where $c[(x1,x2), \theta]$ is the confidence based on the combined data $(x1,x2)$, $c(x1,\theta)$ is that based on the data $x1$, and $L(\theta; x2) = \text{Prob}(X2=x2|\theta)$ is the likelihood based on the data $x2$. This leads to

$$c(x, \theta) \propto c0(\theta)L(\theta; x),$$

where $c0(\theta) \propto c(x, \theta)/L(\theta; x)$ is the induced prior confidence without the data, and $L(\theta; x)$ the likelihood.

The confidence density $c(x, \theta)$ and induced prior $c0(\theta)$ correspond to the Bayesian posterior $\text{Prob}(\theta|x)$ and prior $\text{Prob}(\theta)$, respectively. Thus, the confidence can be obtained by using the Bayes rule (1) under the induced prior confidence. So, confidence is the frequentist alternative to Bayesian logical probability. However, confidence is derived without presuming a prior, whereas Bayesian posterior is based on the prior. But confidence and therefore fiducial probability is not necessarily a Kolmogorov probability, so that we use $c(x, \theta)$ to represent the confidence even though it plays the same role as the Bayesian logical probability. In the Bayesian approach as long as a prior is proper Kolmogorov probability, the posterior is always proper. However, induced prior $c0(\theta)$ is often improper, the resulting confidence may not be a proper probability. Just as a Bayesian posterior contains a wealth of information for any type of Bayesian inference, a confidence density contains a wealth of information for constructing almost all types of frequentist inferences on fixed parameter $\theta$ (Xie and Singh, 2013). For notational convenience, we shall sometimes use Bayesian logical (posterior) probability $\text{Prob}(\theta|x)$ under a prior $c0(\theta)$ for the confidence density $c(x,\theta)$ in the confidence approach.

To summarize, confidence and likelihood are fundamental statistical concepts, currently known to have distinct technical interpretation and usage. Confidence is a meaningful concept of uncertainty within the context of confidence-interval procedure, while likelihood has been used predominantly as a tool for statistical modelling and inference given observed data (Pawitan, 2001; Lee et al, 2017). Pawitan and Lee (2020a) showed that confidence is an extended likelihood, thus giving a much closer correspondence between the two concepts. This result gives the confidence concept an external meaning outside the confidence-interval context, and the extended likelihood theory gives a clear way to update or combine confidence information. On the other hand, the confidence connection gives the extended likelihood direct access to the frequentist confidence interpretation, an objective certification not directly available to the classical likelihood. This implies that inferences from the extended likelihood have the same logical status as confidence interpretations, thus simplifying the terminology in the inference of random parameters.

**Confidence resolution of induction problem**

Let E be the proposition "The sun rises tomorrow" and G be that "it rises forever." Lee (2020) derived a confidence density for the sunrise problem using Pawitan's (2003, Chapter5) right-side P-value, leading to a logical probability

$$P(E| \text{n days of consecutive sunrises}) = 1,$$

so that

$$P(G| \text{n days of consecutive sunrises}) = 1.$$

This allows the realization of complete confidence even with n = 1. Furthermore, P(G|no sunrise at least in one day)=0.

Russell (1912) illustrated induction problem, "Domestic animals expect food when they see the person who usually feeds them. We know that all these rather crude expectations of uniformity are liable to be misleading. The man who has fed the chicken everyday throughout its life at last wrings its neck instead, showing that more refined views as to the uniformity of nature would have been useful to the chicken." Regardless of the number of observations, Hume (1748) would even argue that we cannot claim it is "more probable", since this still requires the assumption that the past predicts the future. Let G be a general proposition that a specific event E occurs always. Let $X_i = 1$ if E occurs at the ith independent observation or experiment and = 0 otherwise. The long-run frequency of $X_i = 1$ is $\theta$. Provided $T_n = X_1+…+X_n = n$, we can claim the uniformity of nature that the event E occurs always with complete confidence. There seems no reason to believe non-uniformity Prob(G)=0 a priori.

In response to the skepticism raised by Hume (1748), Kant (1781) proposed the consideration of the general proposition as absolutely valid Prob(G)=1, a priori, which is otherwise drawn from the dubious inferential inductions. In contrast Bayes (1763) and Laplace (1814) presumed a priori that the general proposition is false Prob(G)=0. Thus, Kant's proposal is consistent only if the general proposition is true, whereas the Bayes–Laplace rule is consistent only if the general proposition is false. It is not necessary a priori to presume P(G) = 0 or 1. Jeffreys (1939) presumed Prob(G)=1/2 to rescue the Laplacian formulation of induction. Now we discuss why the confidence approach provides a resolution of induction problem. Lee (2020) demonstrated that the confidence leads to two potential induced priors, specifically, Beta(0,1) and Beta(1,0). Although these priors are not proper probability, having an infinity measure, they allow a reasonable interpretation. For example, the Beta(1,0) prior indicates that only one success is observed a priori. Thus, if we observe all successes, it is legitimate to attain 100% confidence on $\theta = 1$. However, even if we observe all the failures, we can never attain 100% confidence on θ = 0 because of the success a priori. The Beta(0,1) prior exhibits the contrasting property.

Through deduction, one can achieve complete confidence regarding a particular proposition

$$\text{Prob}(E|G) = 1,$$

provided that the general proposition G is true, Prob(G)=1. Through induction, we can have 0 ≤ Prob(G|E) ≤ 1. Under Bayes-Laplace (Kant) formulation Prob(G|E)= Prob(G)= 0 (Prob(G|E)= Prob(G)= 1) for any evidence E because of presumption Prob(G)=0 (Prob(G)=1) a priori, whereas under Jeffrey's resolution P(G|E) > 0 because he presume Prob(G)=1/2 but cannot reach one (complete confidence) with finite evidence. However, the confidence resolution implies surprisingly that it is legitimate to claim P(G|E) = 1, i.e. one can attain complete confidence regarding the general proposition in finite samples. Confidence approach interprets such a complete confidence as a consistent sample dependent frequentist estimator of the unknown logical probability P(U = 1) = P(I(G) = 1). The estimator becomes more accurate as evidence grows. Lee (2020) showed its theoretical consistency.

To confirm the validity of the general relativity theory, the observational evidence of light bending was obtained in 1919 and the astrophysical measurement of the gravitational redshift was obtained in 1925. Thus, a new theory was confirmed based on a few observations. Then, our resolution shows that it is legitimate to predict the future uniformly with complete confidence unless the general relativity theory stops to hold in the future. Such an inductive reasoning is theoretically consistent and therefore rational. Via induction based on finite data, we can complete confidence that the sun rises forever. (Of course, in physics, the sun runs out of energy, and the solar system vanishes eventually, but here we are discussing only our logical-mathematical confidence given some evidence.) To establish the general proposition from the particular instances by means of induction, scientists do not need to review all the instances but to establish a scientific theory pertaining to the generation of the instances. If one drops an apple, one can be sure that it will fall unless the Newtonian laws suddenly stops to hold. Indeed, it is induction, as we have seen, to allow such uniformity, so that it is the glory of both science and philosophy.

## Extended likelihood as objective logical probability

The likelihood is an uncontroversial technical element that is acceptable to all schools of statistics, but its direct use for inference is controversial (Pawitan 2004). Fisher (1973) recognized the two logical levels of uncertainty, one captured by probability and the other by his classical likelihood, correspondingly two levels of rational thinking. They are not meant to be in competition with each other, as classical likelihood is weaker than logical probability. For example, unlike Bayesian logical probability, the classical likelihood values do not allow an objective frequency-based calibration. Fisher (1930) proposed "fiducial probability" as an alternative to Bayesian logical probability. However, controversies arise as fiducial probability is not necessarily Kolmogorov probability, so that it has been abandoned. Recently, in statistical literature it is appearing as confidence. We claim that the confidence is objective because it can be obtained without assuming a subjective prior. Pawitan and Lee (2020a) showed the confidence is indeed an extended likelihood for unobserved random variable T

$$\text{Prob}(T = I(E) = 1), \qquad (4)$$

which can be viewed as a betting quotient of an event (or proposition) E. Ramsey (1931) and de Finetti (1974) proved that it is coherent as long as it satisfies probability laws. But their definition of logical probability (4) is subjective because people with the same data are allowed to have different logical probabilities. With confidence approach, the confidence (4) is an objective extended-likelihood value, which does not depend upon subjective priors. It can be an objective betting quotient unless there is a relevant subset (Fisher, 1958).

**Postulate of ignorance**

In the sunrise problem Laplace adopted the principle of insufficient reason to justify the use of the uniform prior. We see that it implicitly presumes Prob(G)=0, so that it cannot be an ignorant prior. Fisher (1958) noted that the postulate of ignorance is very important in developing inductive methods. However, he refused to make any axiomatic prior probability. Thus, his fiducial probability (and therefore confidence) does not necessarily satisfy properties of Kolmogorov's probability (Schweder and Hjort, 2016), whereas Bayesian logical probability does if the prior is a proper probability.

According to Gödel (1931), even in a mathematical deductive system, there always exists a proposition G that can be neither proved nor disproved. Thus, mathematics itself also cannot avoid uncertainty. Turing (1936) reformulated Gödel's 1931 results, replacing Gödel's universal arithmetic-based formal language with a simple hypothetical devices known as Turing machine, which is capable of performing any conceivable computation. Turing machine is realized as modern computer. To rephrase Gödel's 1931 problem: can a computer determine whether an arbitrary proposition can be proven or not? Turing (1936) showed that the answer is no. He proved that it is not possible to decide whether a Turing machine will ever halt to return the answer. Uncertainty is also unavoidable in many different ways even in computing. In summary, there is a mathematical proposition G which cannot be proven even if they are true. Likewise, a computer cannot to tell us whether the proposition G is solvable in finite time or not. Thus, even in mathematics and computer sciences there is always a proposition G, whose truthfulness is unknown:

$$T=I(G)=1 \text{ (G is true) or 0 (G is not true).}$$

Thus, even though Prob(G) is either 0 or 1, but it is not possible to know Prob(G) a priori. Human may not perceive the true general laws, but it does not mean Prob(G)=0. Furthermore, to presume Prob(G)=0 a priori does not imply ignorance neither. However, to postulate an ignorance, namely the principle of insufficient reason, Bayes and Laplace used the uniform prior, which turns out to presume Prob(G)=0. We also see Jeffreys's (1939) ignorant prior and Bernardo's (1979) of objective Bayesian schools reference prior presume Prob(G)=0. This is as a strong presumption as Kant's Prob(G)=1. In his rejection of inductive reasoning, Popper also presumed Prob(G)=0, so that in consequence he believed the falsification is the only way to conduct scientific inference. In this formulation Popper's view on inductive reasoning was as extreme as Kant's. We can confirm or falsify general proposition based on evidence, which can be turned out to be wrong later according to future evidence. But that is what our human can do confidently (corroborate) in building our knowledge.

In the 16$^{th}$ century, Michel de Montaigne was most famously known for his skeptical remark, "What do I know?" He was the one who thought deeply about ignorance and concluded that it could not be ended with period. If we are saying we do not know something, then it cannot be an ignorance because we know what is unknown. The best way to represent the ignorance would be to do nothing about it, following Wittgenstein (1921) "What we cannot speak about we must pass over in silence." To represent ignorance by specifying Prob(G)=0 is not ignorance at all, i.e. it is knowledge that G is known to be false a priori.

Deductive logic is based on the knowledge or assumption that certain general propositions or axioms are true. Thus, its conclusions, any derivable statements, are true without any uncertainty. However, inductive reasoning mainly concerns handling of uncertainties, caused by lack of information (ignorance) or limited data. For inductive reasoning, Bayesian school also uses deductive logic by presuming the prior on Prob(G) (like making axioms in mathematics). Whereas, with the confidence or extended likelihood approach, we do nothing on what we do not know; this, we believe, is the most important requirement in developing inductive reasoning. Popper was against the Bayesian logical probability approach, so the non-Bayesian approach here can help realize some of his visions.

## Confirmation problems

We now discuss the difficulties arising in confirming propositions via two well-known fallacies – the conjunction and prosecutor's fallacies. They highlight two distinct modes of reasoning, one captured by logical probability and the other by (classical) likelihood. Our seemingly irrational behavior is due to a decision making based on the likelihood. Thus, from the likelihood perspective, we are still behaving rationally. Recognizing these two modes may lead to better understanding and assessment of our decisions. The difficulties vindicate Popper's criticism of probability-based induction. Also known as Linda problem, the conjunction fallacy originated from Tversky and Kahneman (1983):

> *Linda is 31 years old, single, outspoken, and very bright. She majored in philosophy. As a student, she was deeply concerned with issues of discrimination and social justice, and also participated in anti-nuclear demonstrations. Which is more probable?*
>
>     H1: Linda is a bank teller.
>     H2: Linda is a bank teller and is active in the feminist movement.

Because H2 is a conjunction of two events (hypotheses), it always has lower probability than H1. Yet, from Kahneman (2011, page 158): "About 85% to 90% of undergraduates at several major universities chose the second option, contrary to logic. Remarkably, the *sinners* [our emphasis] seemed to have no shame. When I asked my large undergraduate class in some indignation, 'Do you realize that you have violated an elementary logical rule?' someone in the back row shouted, 'So what?'" After seeing the results of their empirical studies he wrote: "I quickly called Amos [Tversky] in great excitement to tell him what we had found: we had pitted logic against representativeness, and representativeness had won!"

Using probability-based reasoning for Linda problem we are seemingly forced to prefer H1 over H2 *regardless of the data*. This actually feels unnatural: in science it is more reasonable to assume that scientists will formulate and test the strongest hypothesis that is supported by the data, not the safest. This actually corresponds to a seemingly paradoxical Popperian view that, among competing hypotheses, one should in fact adopt the least probable hypothesis that is supported by data. The qualifier "supported by data" in practice of course requires a statistical test. The safest hypothesis can be the blandest, the one with the highest probability of being correct, but has the weakest power to explain the data. 'A feminist bank teller' is 'a bank teller' for sure, but the feminist element makes it a more interesting hypothesis with more explanatory power than the bland 'bank teller'.

We could also add other boring hypotheses such as 'H3: Linda is a woman' or even 'H4: Linda is human', which would have higher probabilities than 'H1: Linda is a bank teller'. Is it 'rational' to prefer H4 when there is enough information pointing to H2? The preference of H2 over H1 is an indication that -- for the people making such judgement – there is enough information pointing to H2. On the other hand, it is not reasonable either to choose 'H5: Linda is a widowed feminist bank teller' as there is nothing in the data supporting the widowed status. So, rationally, the best hypothesis is the strongest hypothesis that is supported by the data; this relates to a notion of optimality, and much hypothesis tests based on the likelihood ratios have been developed to establish the optimality of likelihood-based inference, e.g. Neyman and Pearson (1933).

**Bayesian reasoning in the conjunction fallacy**

The conjunction fallacy has been used as an example of a defect in human reasoning. Despite extensive inquiry, however, the attempt to provide a satisfactory account of the phenomenon has proved challenging. Bayesian confirmation theory has been developed. Inductive logic may be seen as the study of how data (evidence) affect the probability of a proposition H. From Laplace (1814), the posterior Prob(H|data) is considered as an appropriate formalization of the basic inductive logical relationship between evidence and proposition. However, this could lead to counterintuitive consequences and conceptual contradictions (Popper, 1959). There is a fundamental distinction between the notions of logical probability (firmness) and its increase in a proposition H in the light of evidence. Thus, the posterior (logical probability) could be taken as accounting for the former concept, but not the latter (Carnap, 1962). In fact, the degrees of belief on H may increase as an effect of evidence and still remain relatively low (for example, because the disease of interest is very rare). The term "confirmation" has been used in the epistemology and philosophy of science whenever the observational data (evidence) support scientific proposition, meaning in terms of Carnap's increase in firmness brought by data to H. Many Bayesian confirmation measures have been proposed. As an example we consider Carnap's (1950) degree of confirmation of proposition H by the data (evidence)

$$C(H, \text{Data}) = \text{Prob}(H|\text{Data}) - \text{Prob}(H), \qquad (3)$$

which can be positive unless Prob(H) is either 0 or 1. Crupi et al. (2008) derived somewhat complicated conditions under which all confirmation measures satisfy

$$C(H2, \text{Data}) \geq C(H1, \text{Data}).$$

Thus, increase of the probability of H2 by the data can be greater than that of H1. However, a difficulty in Bayesian confirmation is again how to choose the prior, so that it seems arbitrary and complicated conditions are necessary for confirmation.

**Likelihood reasoning in the conjunction fallacy**

The literature on the conjunction fallacy unfortunately does not make a distinction between 'probability' and 'likelihood'. Consider the first part of the description of Linda problem (Linda's characteristics) up to the question as 'Data', and the two statements about her as hypotheses H1 and H2. Then the assessment of H1 and H2 can be either probability-based or likelihood-based. Mixing them up generates the apparent fallacy and confusion.

The likelihood-based reasoning is based on comparing the classical likelihoods of Fisher (1921)

$$L1 = L(H1) = L(\text{Linda is a bank teller}) \equiv \text{Prob}(\text{Data}|H1)$$
$$L2 = L(H2) = L(\text{Linda is a feminist bank teller}) \equiv \text{Prob}(\text{Data}|H2),$$

while the probability-based reasoning is based on comparing the logical probabilities

$$P1 = \text{Prob}(\text{Linda is a bank teller}|\text{Data}) = \text{Prob}(H1|\text{Data})$$
$$P2 = \text{Prob}(\text{Linda is a feminist bank teller}|\text{Data}) = \text{Prob}(H2|\text{Data}).$$

Now, as probabilities, it is always the case that $P2 \leq P1$. But as likelihoods, there is no guarantee at all that $L2 \leq L1$, because likelihood is not a probability of hypothesis.

In Linda problem it is possible that, *when given the description*, the study participants are intuitively making their judgement between feminist vs non-feminist alternatives, thus actually preferring H2 over an unstated but more natural competing hypothesis of non-feminist bank teller. Is the likelihood-based reasoning still consistent with the choice of H2 over H1? Thus, suppose that we have the complementary hypotheses

H2: Linda is a feminist bank teller vs
H3: Linda is a non-feminist bank teller,

and further assume that Linda is a typical female bank teller, i.e. not specially selected, so we can logically compute the necessary probabilities. In likelihood terms, the preference of H2 over H3 is the judgement that

$$L2 = \text{Prob}(\text{Data}|H2) \geq L3 = \text{Prob}(\text{Data}|H3).$$

First note that H1 = {H2 or H3}, i.e. a bank teller is either a feminist or a non-feminist. Among the female bank tellers, let us denote the proportion of feminists as p and the proportion of non-feminists $(1 - p)$. Then, with some probability calculations we have

$$\begin{aligned}
L1 \equiv L(H1) &= \text{Prob}(\text{Data}|H1) \\
&= \text{Prob}(\text{Data}|H2 \text{ or } H3) \\
&= p \times L2 + (1 - p) \times L3 \\
&= L2 - (L2 - L3) \times (1 - p) \\
&\leq L2,
\end{aligned}$$

because the term $(L2 - L3) \times (1 - p) \geq 0$ for any value of p between 0 and 1. Hence the composite hypothesis H1 has a lower likelihood than the constituent likelihood H2. So, within the likelihood framework, the order of preference between H2 and H3 is consistent with the order between H2 and H1.

This is in stark contrast to logical probability-based reasoning, since we always have Prob(H1|Data) ≥ Prob(H2|Data) regardless of the ordering of Prob(H2|Data) vs Prob(H3|Data). In fact, by taking the 'Data' into account, a great majority of the undergraduates and the homunculus are thinking the opposite: implicitly making the judgement that L2 ≥ L1, hence preferring the second hypothesis. In likelihood approach, likelihood ratio L2/L1 ≥ 1 is used to select H2 between hypotheses H1 and H2.

Which reasoning is better? This is not a simple question. Mathematically it depends on how 'Linda' comes into the picture. If she is randomly selected from the population (which determines the sampling probability of Linda), the extended likelihood is defined (Lee et al., 2017). Then, probability(confidence)-based reasoning is mathematically guaranteed to be better, in the sense that it would produce less error. In such cases, the confidence-based reasoning can be also justified under the extended likelihood framework (Lee et al, 2017). If sampling distribution is proper, the confidence is proper probability.

However, the problem description does not make any explicit statement how Linda was selected. Without the random selection, then in principle there is no definite answer; e.g., in the study Linda could be specially selected from among the feminists, in which case H2 is correct. There are many scientific studies that do not rely on random samples. For example: (i) in clinical trials we randomize subjects into study groups; (ii) in epidemiologic studies subjects are often selected based on their outcomes status, resulting in non-random selection. The (classical) likelihood-based reasoning presumes that we know nothing about how Linda comes to the picture, so the likelihoods are the only available *objective quantities* for inference. By preferring H2, the undergraduates are making this stance implicitly. Is that irrational?

As we have stated above, the largest section of scientific statistical data analysis today is based on the classical likelihood. That is, probabilities about the states of nature are rarely included in the analyses, so the analysis is closer in spirit to the classical likelihood-based reasoning above. It is possible to state the problem more carefully so that the logical probability (Bayesian posterior or frequentist confidence) is a better metric for decision, for example by making explicit that Linda was chosen randomly from among 100 women that fit the description, but elaborating on such a situation is not our purpose here. We simply want to provide an explanation of the conjunction fallacy, which is that many people – including the sophisticated undergraduates and homunculus – appear to use classical likelihood-based reasoning in daily life when sampling probability of Linda is uncertain. Hence the 'conjunction fallacy' is not a fallacy, but the result of a mathematically valid likelihood-based reasoning.

# Prosecutor's fallacy

The so-called prosecutor's fallacy can also be explained as a confusion between probability and likelihood-based reasoning. The application of statistical inference in court has been the subject of serious discussion and debates, especially after the emergence of DNA profiling as part of evidence. The logic of legal concepts such as 'presumed innocence' or 'guilt beyond reasonable doubt' has direct statistical connotations, so the principles apply more generally to any assessment of evidence. As discussed in Gardner-Medwin (2005), three key propositions at issue:

> A: the facts or evidence could have arisen if the defendant is guilty
> B: the facts or evidence could have arisen if the defendant is innocent
> C: the defendant is guilty.

Clearly A and not-B together would imply C, but C does not imply not-B. The latter is obvious if the evidence is weak, i.e. could easily have been found among innocent people. Thus strong beliefs in A and non-B together is a more stringent requirement than a belief in C alone. In fact, for expert witnesses, the presumed-innocence requirement may preclude the assessment of C. Since the categorical truth of these statements is in reality rarely available, the prosecution may have to present extremely small probabilities to establish guilt beyond reasonable doubt.

Those probabilities of A and B are in fact likelihoods of guilt:

> L1 = L(Defendant innocent) = Prob(Evidence|Defendant innocent)
> L2 = L(Defendant guilty) = Prob(Evidence|Defendant guilty)

while the probabilities of guilt are

> P1 = Prob(Defendant innocent|Evidence)
> P2 = Prob(Defendant guilty|Evidence) .

Thus in a typical court proceeding, what is computed is the likelihood L1. The so-called prosecutor's fallacy is to misrepresent L1 as P1. Putting aside any technical issues in its computation, suppose L1 is very low, say $10^{-8}$. And suppose further that the probability of DNA matching is one if the defendant is guilty, i.e., L2 = Prob(Evidence|Defendant guilty) = 1. So the likelihood-based reasoning leads to a likelihood ratio

$$L2/L1 = 10^8,$$

Which provides the joint assessment of (A and non-B) propositions. In using L1 and L2 directly without prior probabilities of guilt, the prosecutor is relying on a *valid likelihood-based reasoning. It is not a fallacy.*

One may of course argue that probability-based reasoning based on P2/P1 is better, but this will require the establishment of prior probabilities of guilt that can be agreed by all parties. One can easily imagine the contentious arguments on settling the prior probabilities; for example, is it reasonable to presume that the defendant is a random sample from the general population? How do we abide by the presumed innocence requirement? However, again our purpose here simply to point out that the prosecutor's argument in fact does not have to rely on the logical probabilities (confidences), but on a valid (classical) likelihood-based reasoning, hence avoiding the fallacy. Unfortunately in layman language, 'probability' and 'likelihood' are interchangeable as expressions of uncertainty, thus confusing the two valid modes of reasoning and making it impossible for the prosecutor to avoid the fallacy.

## Traces of likelihood-based learning in infants

Which is more natural: probability-based or likelihood-based reasoning? Gweon et al. (2010) presented a fascinating series of experiments on inductive learning by infants -- average age 15 months old -- as evidence that likelihood-based learning is perhaps hard-wired in our brains. The story is delightfully told in Laura Schulz's Ted Talk: How do babies' "logical minds" work? [https://www.ted.com/talks/laura_schulz_the_surprisingly_logical_minds_of_babies](https://www.ted.com/talks/laura_schulz_the_surprisingly_logical_minds_of_babies).

Here we only highlight their key experiment: the babies were presented with *3 squeaky blue balls* taken from a large opaque box. The box-wall facing the babies had a clearly visible picture indicating its content. Two scenarios were performed:

> **Scenario 1**: the picture showed mostly blue balls and some yellow balls;
> **Scenario 2**: the picture showed mostly yellow balls with some blue balls.

(The ratio is 3:1 in each case, but as far as the babies were concerned we suppose the exact number did not really matter.) The blue balls were taken one at a time, each time shown to the babies that they squeaked. Then babies were given a *single yellow ball*; the question is would they attempt to squeak it? What reasoning or inference method do they use?

First we can agree that the only Data available to the babies are {3 squeaky blue balls}. To use the probability-based reasoning the babies would have to come with the Bayesian posterior probability

$$\text{Prob(yellow balls are squeaky|Data)},$$

which would of course require the prior probability before seeing the data and the use of Bayes's formula to compute the posterior. Presumably, if the baby judges the probability to be high enough, then they will try to squeak the given yellow ball. However, the necessary calculation looks too difficult for most babies we know. The likelihood-based reasoning would require the babies to assess

$$L(\text{yellow balls are squeaky}) = \text{Prob}(\text{Data}|\text{yellow balls are squeaky}).$$

This is perhaps not obvious either, as there is never any direct evidence of the squeakiness of yellow balls; so, the inference must somehow come from an inductive generalization.

Gweon et al (2010) described a model involving 4 hypotheses leading to predictions based on likelihood reasoning, but here we shall construct a simpler thought process. On seeing 3 squeaky blue balls, the babies were implicitly assessing these 2 hypotheses:

> H1: the sample is randomly selected from all the balls
> H2: the sample is not random, but selectively taken only from squeaky blue balls

Furthermore, with inductive generalization, when a sample was judged random then the properties of the sample would generalize; for instance, here, squeakiness then applies to all balls, hence to the yellow balls. And vice versa, when a sample was not random, then the properties would not generalize.

On observing 3 squeaky blue balls, the likelihoods of the hypotheses are now computable. For Scenario 1 (mostly blue balls):

$$L_1 = \text{Prob}(\text{Data}|H_1) \sim \text{high}$$
$$L_2 = \text{Prob}(\text{Data}|H_2) \sim \text{high}.$$

Presumably the babies did not use the exact values, but used only visual clues ($L_1/L_2 \sim 1$) to conclude that there was no reason to reject the random sampling hypothesis H1. So the squeaky property generalized to yellow balls, and the babies were predicted to squeak the yellow balls. On the other hand, for Scenario 2 (mostly yellow balls):

$$L_1 = \text{Prob}(\text{Data}|H_1) \sim \text{low}$$
$$L_2 = \text{Prob}(\text{Data}|H_2) \sim \text{high}.$$

Again using only visual clues ($L_1/L_2 \ll 1$) to judge the hypothesis, babies would reject the random sampling hypothesis H1, hence not generalize the squeaky property. So they were predicted not to squeak the yellow balls. The results confirmed these predictions, or we could say the likelihood-based reasoning explains the experimental results, providing evidence of elementary use of likelihood-based reasoning in infants.

## Discussion

Popper's views on scientific inference seems derived from and more suitable for the hard sciences, particularly physics. Indeed, in his *Logic of Scientific Discovery,* he had a whole chapter on the quantum theory, while his propensity theory of probability applies naturally to physical phenomena. He somewhat downgraded the role of experience, empiricism and induction in the scientific discovery process; for him, all observations are 'theory laden,' i.e. the theory comes before observations. Supposedly, without any theory to begin with, how would anyone even know what observations to collect? According to a perhaps apocryphal story, at the start of his course on philosophy of science he liked to tell his students, 'Go ahead and observe!' Then he would just sit and wait; this was meant to show that without any theory there was nothing to observe. Ironically, one thing all scientists must know is that it is dangerous to generalize from single episodes: how much can we say about the logic of scientific discovery from an observationally barren class-room?

It is instructive to contrast physics with truly empirical sciences such as economics or medicine, where observations are collected all the time, with a purpose for sure, but mostly without any theory. For instance, consider the cancer registration, which, in most countries, is mandated by law. Observations such as cancer incidence are collected without any theory in mind; they are simply done for the purpose of monitoring of disease burden and health planning. But a doctor reading through the cancer records may notice, that certain professions like, for instance, chimney sweeps have much higher rates of scrotal cancer, while miners have higher rates of lung cancer. One may conjecture theories what those risks are, but the original observations that lead to the theories were not themselves ´theory laden'. Endless hypotheses can be formed by going through cancer registry data that cannot be conjectured by pure thinking. In recent medical genetics research, the most successful approach is the hypothesis-free genome-wide studies. This era came after the much lamented fruitless decades of the so-called candidate-gene approach with 'theory-laden' observations. Popper's emphasis on the hard-science-based scientific discovery had created an unnecessarily hard demarcation between deductive and inductive logic. In this paper we describe statistical ideas that are to a large extent Popperian, but also contains the logical elements of inductivism as captured by the Fisherian and Bayesian statistics. Specifically, these are represented by the likelihood and confidence concepts.

With a regular use of the inverse probability method of Laplace, 19th-century statistics was largely Bayesian. Fisher (1930) criticized the use of inverse probability method due to its arbitrary presumption of prior probability, but his own solution preserved the logico-inductive content of probability. There is a virtual consensus regarding the use of probability for statistical modeling, but we have yet to reach that for its interpretation and philosophical aspects. Mathematically, Kolmogorov's axiomatic foundation puts probability as the legitimate child of the mature measure theory. Kolmogorov's probability is naturally interpreted as long-run frequency. Thus, in most statistical frequentist textbooks, probability is said to be meaningless for specific single events such as Donald Trump's impeachment or re-election. But people do bet on such specific events, which can only mean that they do have a logical probability that is not a long-term variety. The reasoning requires a logical probability that applies to specific single events. Moreover, since different people have different beliefs and temperaments, they may have different subjective logical probabilities for the same event even though they share the common information (evidence).

We have described alternative measures of uncertainty including classical likelihood and confidence, and highlighted the differences with Bayesian logical probability. To a large extent this is done in the Popperian spirit of non-probabilistic corroboration. But with confidence and extended likelihood, we are preserving the logico-inductive spirit of Fisher. Likelihood and probability are bread-and-butter concepts in routine statistical analyses of scientific data. What do we gain from distinguishing the (logical) probability- from the (inductive) likelihood-based reasoning? Primarily, clarifying the meaning of the terminologies will also clarify our thinking, thereby reducing unnecessary confusions. For example, we believe there is no need to call the conjunction fallacy a fallacy, and to accuse ourselves as being illogical or irrational, when we are in fact using likelihood-based reasoning. Closing the gap between the technical meaning and layman understanding is always a difficult challenge in the public dissemination of science, but perhaps not hopeless. At the very least, the distinction between probability- vs likelihood-based reasoning should be part of a standard scientific discourse on decision making. The likelihood-based reasoning should be recognized as an objective and rational mode of reasoning. Recently, likelihood has been extended to allow the sense of uncertainty associated with a realized but still unobserved single event, while at the same time avoid potential probability-related paradoxes (Pawitan and Lee, 2017). Extended likelihood and therefore the confidence is not necessarily probability, so that care is necessary when expected utility is computed using confidence (Pawitan and Lee, 2020b).

Even in statistics after 100 years since its introduction, there is still no general consensus on the direct use of likelihood for inference, indicating that it is difficult to give a normative answer. In statistics literature we can point to Edwards (1992) and Royall (1997) as proponents of this mode of likelihood inference, although we must add that they do not represent the Fisherian views we state above. Fisher (1973) recognized the two logical levels of uncertainty, whereas in this article we explain four levels, namely, classical likelihood, confidence, logical probability and Kolmogorov mathematical probability. Extended likelihood is proposed for simultaneous inferences of fixed and random unknowns (Lee et al., 2017). Classical likelihood is that for fixed unknowns, whereas confidence is that for binary random unknowns. We hope that these distinction enrich methodological developments in many non-statistical areas.